\pdfoutput=1
\documentclass[10pt]{iopart}

\usepackage{graphicx}
\begin{document}

\title[Interaction of a Quantum Dot With Non-Classical Light]{Peculiarities of Interaction of a Quantum Dot with Non-Classical Light in the Self-Phase Modulation Regime}

\author{S N Balybin$^{1,3,*}$, R V Zakharov$^{1,2}$, O V Tikhonova$^{1,2}$}

\address{$^1$ Faculty of Physics, Lomonosov Moscow State University, 119991, Moscow, Russia}
\address{$^2$ D.V. Skobeltsyn Institute of Nuclear Physics, Lomonosov Moscow State University, 119234, Moscow, Russia}
\address{$^3$ Russian Quantum Center, Skolkovo IC, 121205, Moscow, Russia}

\ead{$^*$ sn.balybin@physics.msu.ru}
\vspace{10pt}

\begin{abstract}
Influence of the self-phase modulation of quantum light on the induced resonant excitation of a semiconductor quantum dot is studied analytically in the case of the Kerr-nonlinearity of the medium. The phase nonlinearity is found to result actually in a resonance detuning specific for each field photon number state. This effect is shown to provide significant decrease of the excitation efficiency accompanied at the same time by more regular excitation dynamics obtained even for initial squeezed vacuum field state. The enhancement of entanglement between semiconductor and field subsystems with growing non-linearity is demonstrated. As a result, the formation of different types of non-Gaussian field states is found with features being analyzed in details.
\end{abstract}


\ioptwocol

\section{Introduction}

Experimental generation of non-classical states of electromagnetic fields opens new possibilities in modern quantum optics and quantum information science. Different non-classical states such as few-photon Fock and coherent states with small mean photon number, biphoton pairs and multiphoton squeezed states of light can be obtained \cite{McKeever2004,Maunz2007,Schnabel2017}. The interaction of semiconductor systems and nanostructures with non-classical field is one of perspective and interesting problems of modern solid state physics and quantum optics and can lead to new fundamental effects \cite{Kasprzak2010,Sete2010,Sete2011}. At the same time different practical applications based on these effects can be developed. In experiment, a solid cavity is usually used and different types of nonlinear effects can be expected \cite{Kasprzak2010} but are mostly ignored in the theoretical treatment. An important aspect of such investigations is a formation of non-Gaussian states of field induced due to nonlinearity of the cavity medium \cite{Yurke1986,Thenabadu2019,Strekalov2016} The physical mechanism of this effect can be understood using a simple model of nonlinear oscillator. In this case the anharmonicity of the potential affects on the quantum field and leads to the possible appearance of non-Gaussianity \cite{Hu2011}. The non-Gaussian states are experimentally demonstrated to be very promising to develop quantum information protocols \cite{Kirchmair2013,Grimm2020}. For this reason the generation and study of these states are strongly required. At the same time, the formation of such states during the interaction with solid nano-system can significantly change the coupling strength and the dynamics of the process. 

It should be noticed that even the analysis of the Kerr self-phase modulation itself is often restricted to consideration of only coherent states of light and mainly in the limit of extremely small nonlinearity \cite{Milburn1986,Lando2019,Korolkova2001,Dodonov2003}. At the same time the role of the Kerr-nonlinearity during the interaction of light with quantum atomic and semiconductor systems seems to be poorly studied. 

In this paper, we investigate the excitation of a semiconductor quantum dot by non-classical electromagnetic field in a nonlinear cavity and the influence of the Kerr self-phase modulation effect on the interaction process is examined. Different input field states and degrees of nonlinearity are considered and the time-dependent behavior of the quantum dot excitation is analyzed. In addition, the evolution and features of the quantum field state changing during the interaction under self-phase modulation are studied using the Wigner function formalism \cite{Kenfack2004, Schleich2008} and a “quantum carpet" method \cite{Kazemi2013, Leibscher2009} providing deep insight to the considered physics. Deformation of the field wave packet and quadrature squeezing are analyzed in details. 

\section{Theoretical approach}
We consider the interaction of a single photon mode with a semiconductor quantum dot in a solid nonlinear cavity using Jaynes-Cummings model \cite{Kasprzak2010,Shore1993}. The Hamiltonian of the system in the rotating wave approximation is given by:

\begin{equation}\label{ham}
    H =\hbar \omega_0 \frac{\sigma_z}{2} +\hbar \omega a^\dag a - \hbar \frac{g}{2} (a^\dag)^2 a^2 + \hbar \Omega (a^\dag \sigma_{-} + a \sigma_{+}).
\end{equation}

Here the operator $a^\dag, a$ stand for creation/annihilation photon operators, while the $\sigma$-operators describe the semiconductor quantum dot. The third term describes the degenerate four-wave mixing nonlinear process \cite{Milburn1986,Bajer2002} and is responsible for the self-phase modulation of the field.
The non-stationary Schr\"{o}dinger equation with Hamiltonian (\ref{ham}) can be solved analytically and the obtained solution can be expanded over the free eigenstates of subsystems $|n\rangle$ and $|g\rangle,|e\rangle$ (for photon mode and quantum dot respectively):

\begin{equation}\label{sol_form}
    \psi =\sum_{n=0}^{\infty} \left( a_n |g\rangle |n\rangle \e^{-\frac{i}{\hbar} E_{gn} t }+ b_n |e\rangle |n\rangle \e^{-\frac{i}{\hbar} E_{en} t }\right)
\end{equation}

with energies equal to the sum of correspondent free eigenergies:

\begin{equation}\label{sol_e}
E_{gn} = E_{g}+\hbar\omega(n+1/2), E_{en}= E_{e}+\hbar\omega(n+1/2).
\end{equation}

Further a resonant case is considered. The found quasienergies of dressed states for quantum dot coupled with quantized field are given by: 
\begin{equation}\label{qe}
\gamma_{1,2} = -\frac {g(n-1)^2}{2} \pm \sqrt{ \Omega^2 n + \frac{g^2(n-1)^2}{4}}.
\end{equation}

If initially the quantum dot is prepared in the ground state and the field is supposed to be in some superposition of Fock states $\sum_n C_n |n\rangle$ the probability amplitudes in (\ref{sol_form}) are found in the form

\begin{eqnarray}\label{sol}
\nonumber \frac{a_n(t)}{C_n} =\frac{1}{2} (1-\frac{g(n-1)}{2s_n})  \e^{-i \gamma_1 t} + \frac{1}{2}(1+\frac{g(n-1)}{2s_n})  \e^{-i \gamma_2 t},  \\
\frac{b_{n-1}(t)}{C_n} = \frac{\Omega \sqrt{n}}{2s_n}  \left(  \e^{-i \gamma_1 t} -  \e^{-i \gamma_2 t} \right),
\end{eqnarray}
where $\Omega$ is the vacuum Rabi frequency and the notation $s_n=\sqrt{\Omega^2 n+g^2 (n-1)^2/4}$ is used.

The found solution allows to obtain all the information about the system dynamics. The probability of quantum dot excitation $P(t)$ can be calculated as follows:

\begin{equation}\label{pe}
P(t)=\sum_{n=0}^{\infty} \left| b_n(t) \right|^2.
\end{equation}

For further analysis we consider the field quadrature $x=(a+a^\dag)/\sqrt 2$. Using the time-evolution of the found wave function (\ref{sol_form}) the field wave packet is given by
\begin{equation}\label{sol_carpet}
|\psi (x, t)|^2 = \sum_{n,k} F_n(x) F_k(x) \e^{ i \omega (k-n) t} (a_n a_k^* + b_n b_k^*),
\end{equation}
where $F_n(x) = \langle x | n \rangle$ is pseudocoordinate representation of the Fock state. While the variance of the field quadrature can be calculated as follows:
\begin{equation}
Var[x](t)=\langle\psi(t)|x^2|\psi(t)\rangle - \left|\langle\psi(t)|x|\psi(t)\rangle\right|^2.
\end{equation}

Different non-classical input states are examined and the interaction dynamics is found for coherent light with small mean photon number and squeezed vacuum light. The correspondent field states are presented as an expansion over Fock states \cite{Schleich2008}:

\begin{equation}
    |\alpha\rangle = \sum_{k=0}^{\infty} \e^{-|\alpha|^2/2}\frac{\alpha^k}{\sqrt{k!}}| k\rangle,
\end{equation}

\begin{equation}
    |R\rangle = \frac{1}{\sqrt{\cosh r}} \sum_{k=0}^{\infty} (-\tanh r)^k  \frac{\sqrt{(2k)!}}{2^k k!} |2 k\rangle.
\end{equation}

Here the coherent state $|\alpha\rangle$ is characterized by the parameter $\alpha$ which determines the mean photon number of such state $\langle n \rangle = |\alpha|^2$. For the squeezed vacuum state only even Fock states are populated, $r$ is the squeezing parameter and $R=\exp(r)$ is the squeezing  factor which determines the quadrature  squeezing of this state.

\section{Results and discussions}

\subsection{Dynamics of quantum dot excitation}

\begin{figure}[t]
  \includegraphics[width=\linewidth]{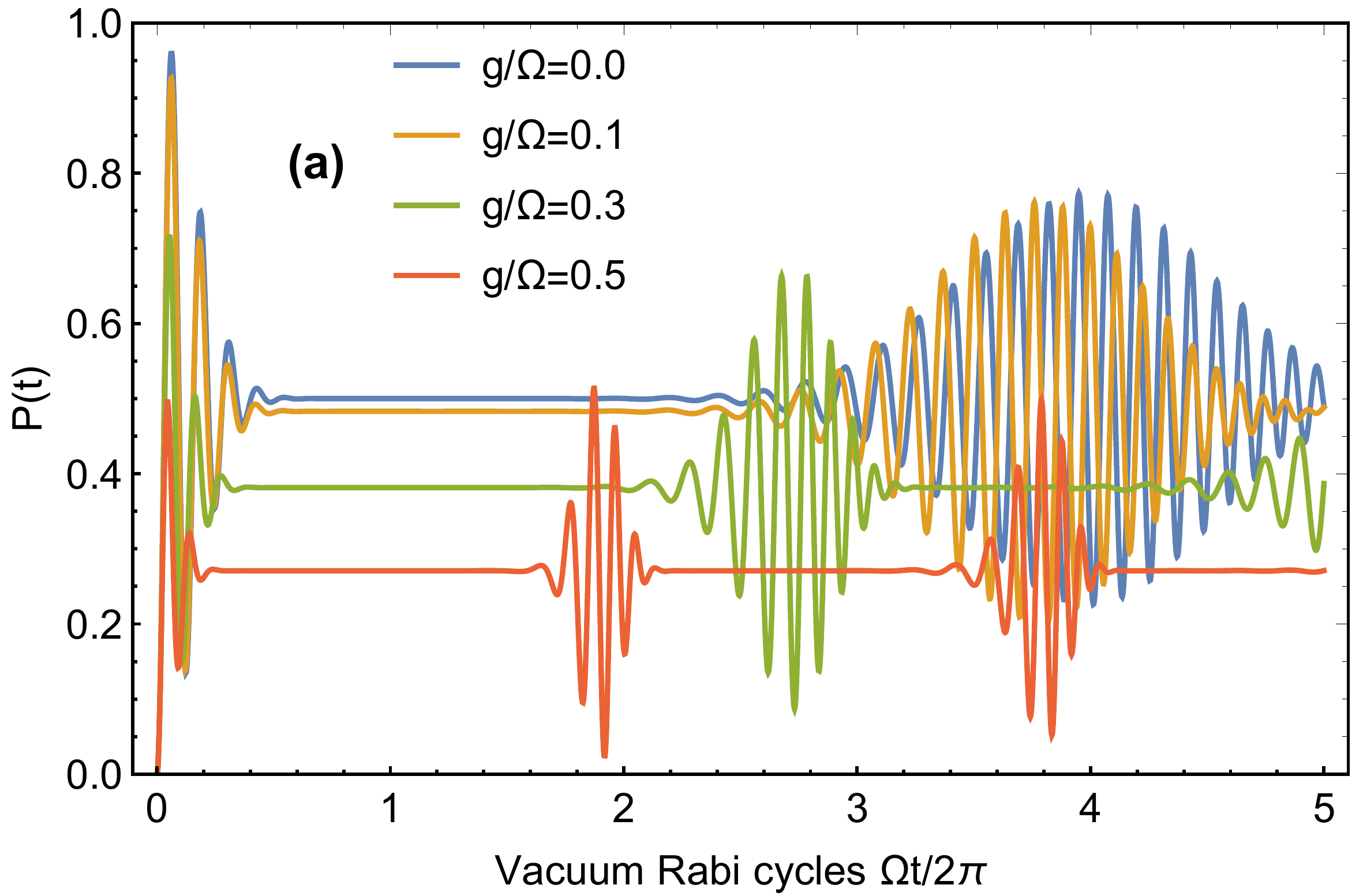}
  \includegraphics[width=\linewidth]{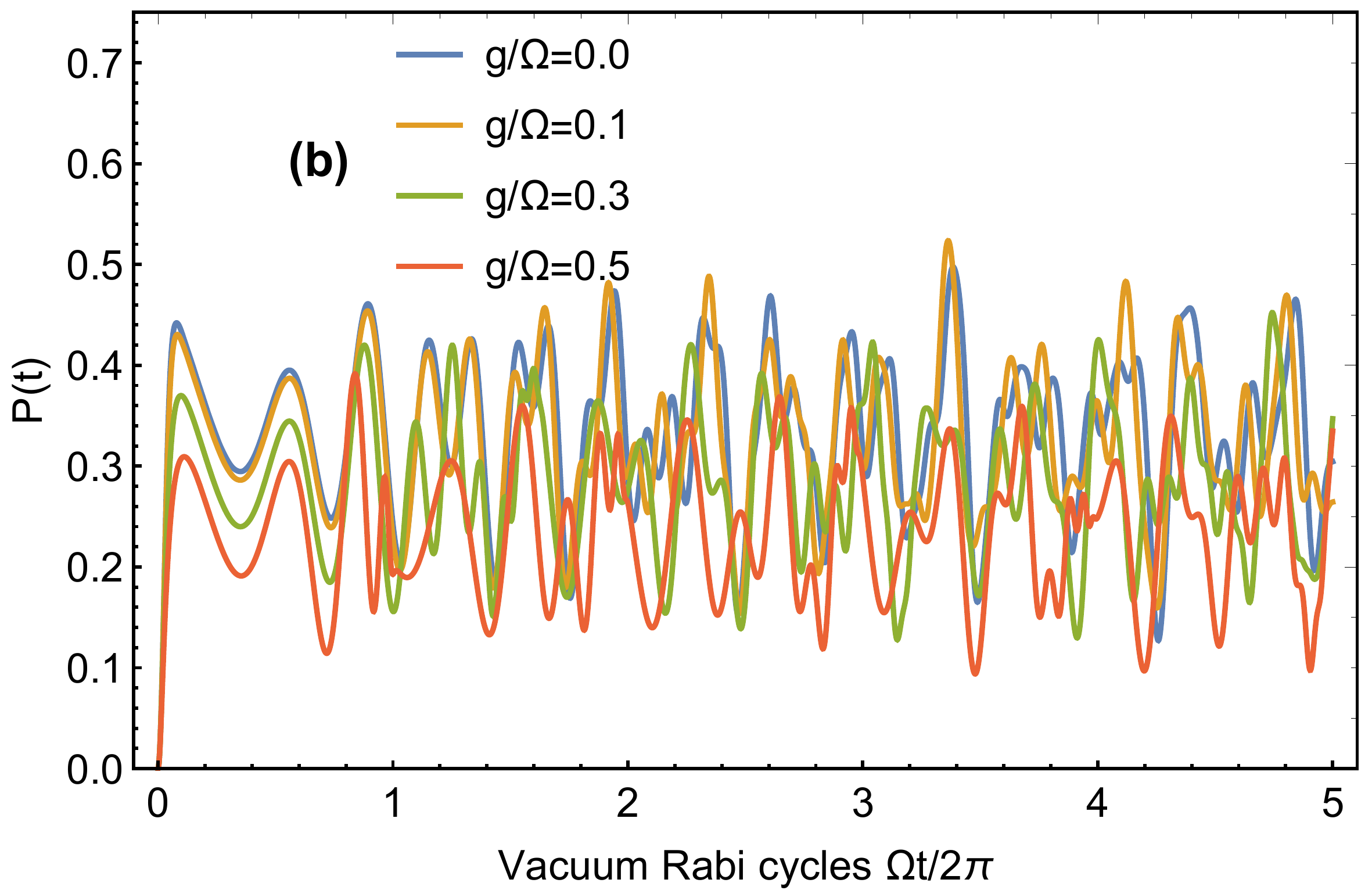}
  \caption{Time-dependent probability of quantum dot excitation obtained for different ratios $g/\Omega$ for the coherent state with $\alpha = 4$ (a) and squeezed vacuum state with $R=6$ (b) of input field.}
  \label{fig1}
\end{figure}

The found time-dependent probability of quantum dot excitation obtained for different degrees of self-phase modulation is presented in Fig.~\ref{fig1} for coherent and squeezed input field states.

Figure \ref{fig1}(a) represents a well-known regimes of collapses and revivals taking place for coherent input field state \cite{Scully1997,Popolitova2019}. It can be seen that for larger nonlinearity the excitation probability is suppressed. The reason for this effect can be understood from the analytical solution (\ref{qe},\ref{sol}) which shows that the self-phase modulation of the field actually results in a detuning from resonance specific for each photon number state. This effect is found to be more pronouced for fields with larger mean number of photons. At the same time the dynamics of the excitation process seems to become more regular since the revivals appear to be shorter in time and the probability of excitation is mostly at the same level determined by the contribution of many oscillating term in (\ref{pe}). Similar behavior is found for squeezed initial field state (see figure \ref{fig1}(b)). Though the regimes of collapses and revivals do not take place in this case, more regular oscillation dynamics is found.

\subsection{The Case of predominant role of nonlinearity}

In the limit of predominant nonlinearity, the coupling between the field and quantum dot appears to be negligibly small and the field state evolution is found to be determined by the self-phase modulation only. In this case the evolution of the photon operators can be found explicitly: 

\begin{equation}
    a(t) = \exp{(i (g \hat n - \omega) t)} a_{in},
\end{equation}

and variance of the field quadrature obtained in the case of initial coherent field state $|\alpha\rangle$ can be calculated analytically:

\begin{eqnarray} \label{var}
     Var[x(t)]=1/2 + |\alpha|^2 \left[ 1-e^{-2 |\alpha|^2(1 - \cos{gt})} \right.\nonumber\\
     -  \left. \cos(2|\alpha|^2\sin{gt}-2\omega t )e^{-2 |\alpha|^2(1 - \cos{gt})}+\right. \nonumber\\
     \left. +\cos( gt+|\alpha|^2\sin{2gt}-2\omega t)e^{-|\alpha|^2(1 - \cos{2gt})}\right].
\end{eqnarray}

\begin{figure*}[t]
  \includegraphics[width=\linewidth]{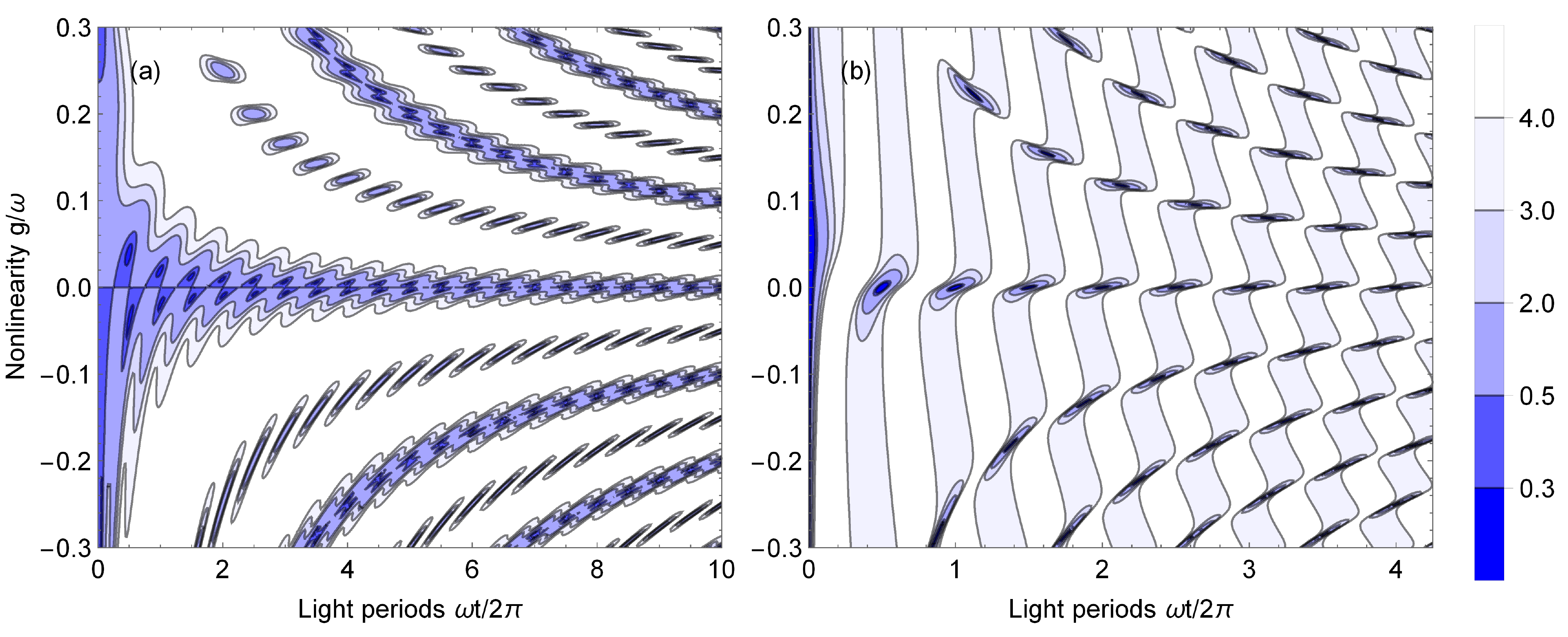}
  \caption{2D plots of the field quadrature variance calculated in dependence on time and for different strengths of nonlinear interaction $g$ for coherent field with $\alpha=2$ (a) and squeezed vacuum  with $R=4$  (b) as initial state.}
  \label{fig2}
\end{figure*}

Being below the shot noise level $1/2$ the obtained variance characterizes squeezing of the field quadrature which consists in significant  decrease of the quantum uncertainty of the field. Figure \ref{fig2} represents 2D plots of the time-dependent variance calculated for different ratios $g/\omega$ for coherent and squeezed vacuum initial field states. 
For coherent initial state the regions of small values of parameter $g$ and time are most interesting when significant squeezing of the field quadrature takes place and its variance becomes less than $1/2$. This regime was already analyzed earlier \cite{Bajer2002}. However  figure \ref{fig2}(a) evidently shows that squeezing is periodically achieved in time and its high degree can be observed for any values of parameter $g$. The realistic values of the ratio $g/\omega$ are significantly smaller than unity though such nonlinearities are rather large in semiconductor media and can be  enhanced in the case of resonance processes. Note that the sign of $g$ is important only in the limit of very low nonlinearity. Thus, the coherent fields are found to be very perspective for squeezing via self-phase modulation. In the case of squeezed vacuum light (see figure \ref{fig2}(b)) the variance of field quadrature is already strongly suppressed initially. And the squeezing is found only periodically reproduced at the level not better than the initial one. As a result for such states there is no possibility to provide significantly higher degree of squeezing due to self-phase modulation. This result is caused by dramatically broad distribution of the squeezed vacuum state over Fock states and strong dephasing of the amplitudes of the individual Fock states  from each other due to self-phase modulation.

\begin{figure}[t]
\centering
\begin{minipage}{.5\textwidth}
  \centering
  \includegraphics[width=\linewidth]{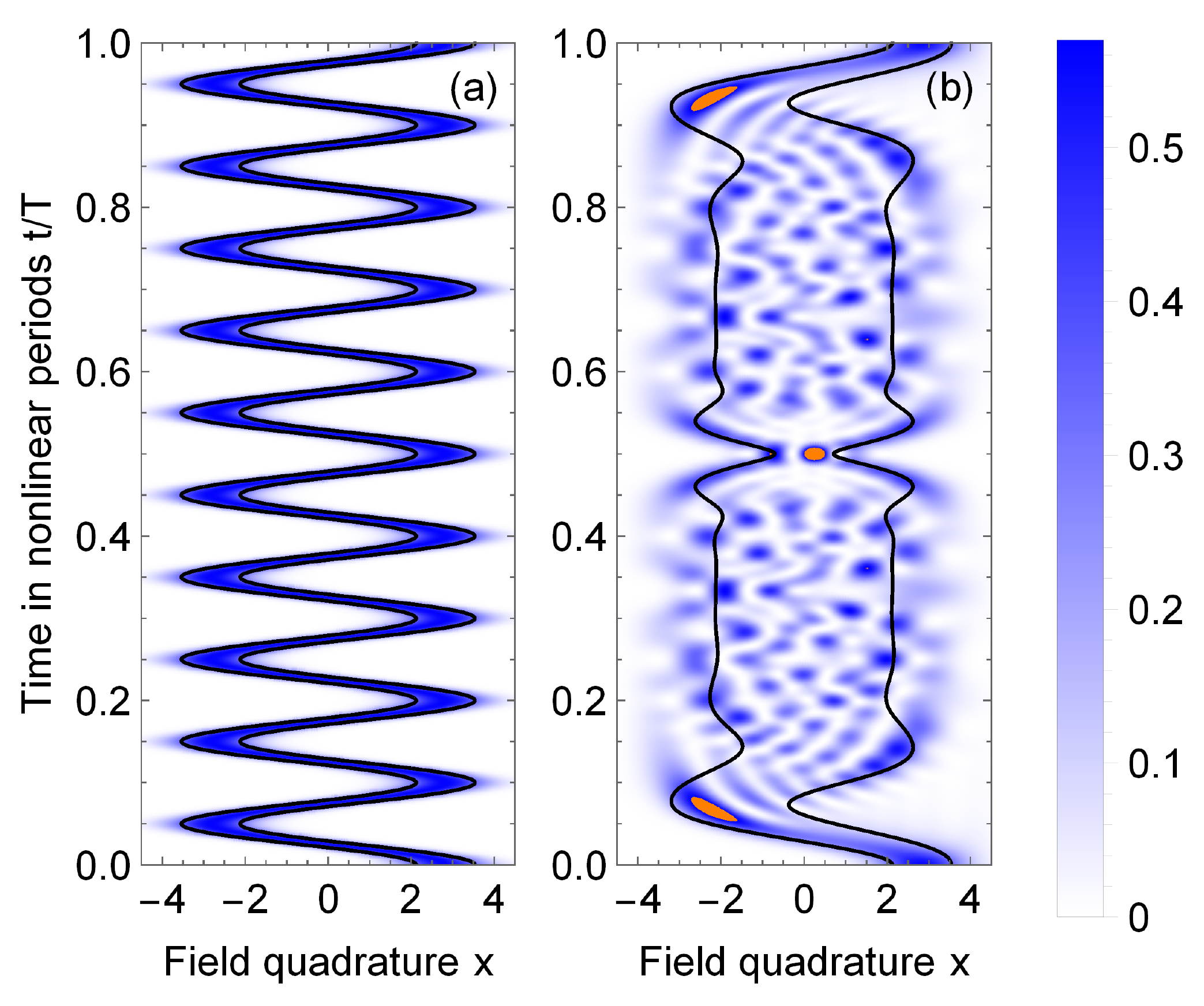}
  \caption{The quantum carpets for initially coherent field with $\alpha=2$ for $g/\omega=0$ (a) and 0.1 (b).}
  \label{fig3}
\end{minipage}
\hspace{10pt}
\begin{minipage}{.5\textwidth}
  \centering
  \includegraphics[width=\linewidth]{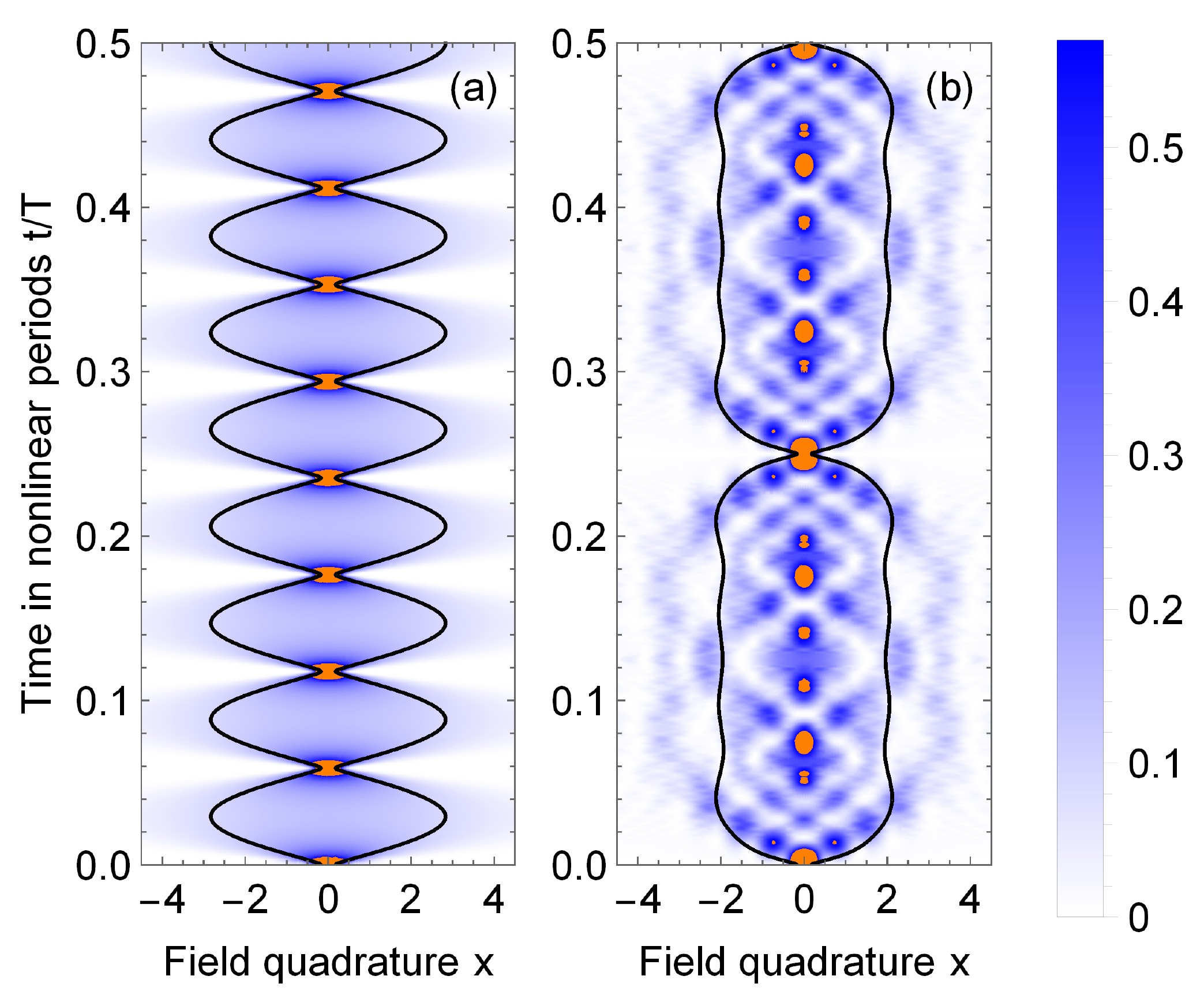}
  \caption{The quantum carpet for initially squeezed vacuum field with $R = 4$ for $g/\omega=0$ (a) and 0.125 (b).}
  \label{fig4}
\end{minipage}
\end{figure}

The data in  figure \ref{fig2} does not give the whole picture of the quantum field dynamics. It should be noticed that the variance gives only an averaged information about time evolution which would be complete for Gaussian states. But for arbitrary state this information cannot describe all features of nonlinear evolution. For this reason it is much more useful to analyze the so-called quantum carpet. It is determined as probability distribution of the quantum field over its quadrature $|\langle x|\psi(t)\rangle|^2$ depending on time (\ref{sol_carpet}). Using the wave function (\ref{sol_form}) we can easily calculate carpets for different initial fields. The results are given in figure \ref{fig3} for initial coherent field state with $\alpha = 2$ and in figure \ref{fig4} for squeezed vacuum light with $R = 4$.
At the left part of each figure the carpets for only free evolution without self-phase modulation are presented. To provide more evident analysis the standard deviation of the $x$-quadrature from its average value $ \sigma[x]=\sqrt{Var[x]}$ is presented at the same time as a thick black line and by the orange color the zones of local squeezing of the field distribution are highlighted.

The interaction-free evolution is characterized by oscillations of the center of mass of the wave packet for coherent state (see  figure \ref{fig3}(a)) and its width for squeezed vacuum state (see  figure \ref{fig4}(a)) with Gaussian shape of the field wave packet being conserved at any instant of time. In the frame of the Wigner function approach such oscillations correspond to the rotation on the phase plane and are usually cancelled out by consideration of the rotating field quadrature. However, the correct time-dependent dynamics of the field wave packet does really include this oscillation behavior.  Moreover, for many practical applications especially in the case of interaction with atoms and nanostructures a certain field quadrature is important for excitation and its magnitude determines the efficiency of this process. For this reason we consider only the electromagnetic field quadrature $x$ and avoid the transformation to the rotating frame.

It is clear that in such interaction-free behavior the mean value and the variance of field quadrature gives full information about the dynamics. But it is not the case for the nonlinear interaction regime. Figure \ref{fig3}(b) and  \ref{fig4}(b) show that for both field states the self-phase modulation induces strong deformation and structuring of initial field wave packet and provides its strong deviation from Gaussian shape. In this case the field features can no longer be fully determined by the average field value and variance. Several local maxima appear on the field probability density leading to the formation of a non-Gaussian field state. The observed structuring of the wave packet seems to be of random character, but a periodic repetition of the field distribution shape in time and periodic total revival of the initial input field wave packet are found and demonstrated. The period of revival is determined by the degree of nonlinearity and is found to be equal to $T=2\pi/g$. Such behaviour can be usually observed for superposition of discrete states (even for infinite number), but the condition of the revival is very sensitive to the dependence of the nonlinear phase on the photon quantum number (see ${\cite{Scully1997}}$ and ref. therein).

In addition to the non-Gaussian character, the formation of several local peaks or even single sharp maximum on the field probability density (highlighted in orange) leads to another important feature of quantum field induced by self-phase modulation. Strong local squeezing of the field quadrature is found in this case, while the total width of the wave packet as well as the variance appears to be rather large. For coherent initial state such behavior can be seen in the vicinity of time $t/T = 0.5$ when the formation of a sharp well-localized wave packet is found. Actually there is another rather narrow side maximum at this time which is not very well seen but establishes a negative field quadrature. Such distribution is found to correspond to formation of a cat state which is a superposition of two coherent states $ (e^{-i \pi/4}|i\alpha\rangle+e^{i \pi/4}|-i\alpha\rangle)/\sqrt{2}$ similar to that found by Yourke for the propagation through the amplitude-dispersive medium \cite{Yurke1986}. The observed local squeezing and formation of one main maximum means that in its vicinity the quantum uncertainty becomes very low and the field quadrature takes an almost exact value. The found focusing of the quantum field in time can be used in many promising practical applications such as supersensitive quantum measurements with noise reduction and selective excitation of atomic and nano-systems.

For squeezed vacuum light local squeezing and formation of sharp peaks on the field probability distribution take place as well, but the value of the field quadrature appears to be close to zero. The observed “focusing” of the electromagnetic field could seem to be nothing more than the reconstruction of the initial wave packet. But it is so only at times that are multiples of the nonlinearity period. In other cases the width of the local maximum appears to be much less in comparison to the initial one, corresponding to dramatically smaller local uncertainty of the electromagnetic field. Presence of the additional poorly pronounced side peaks found at the same time unambiguously indicates a significantly non-Gaussian character of the formed field state. 

\begin{figure}
\centering
\includegraphics[width=\linewidth]{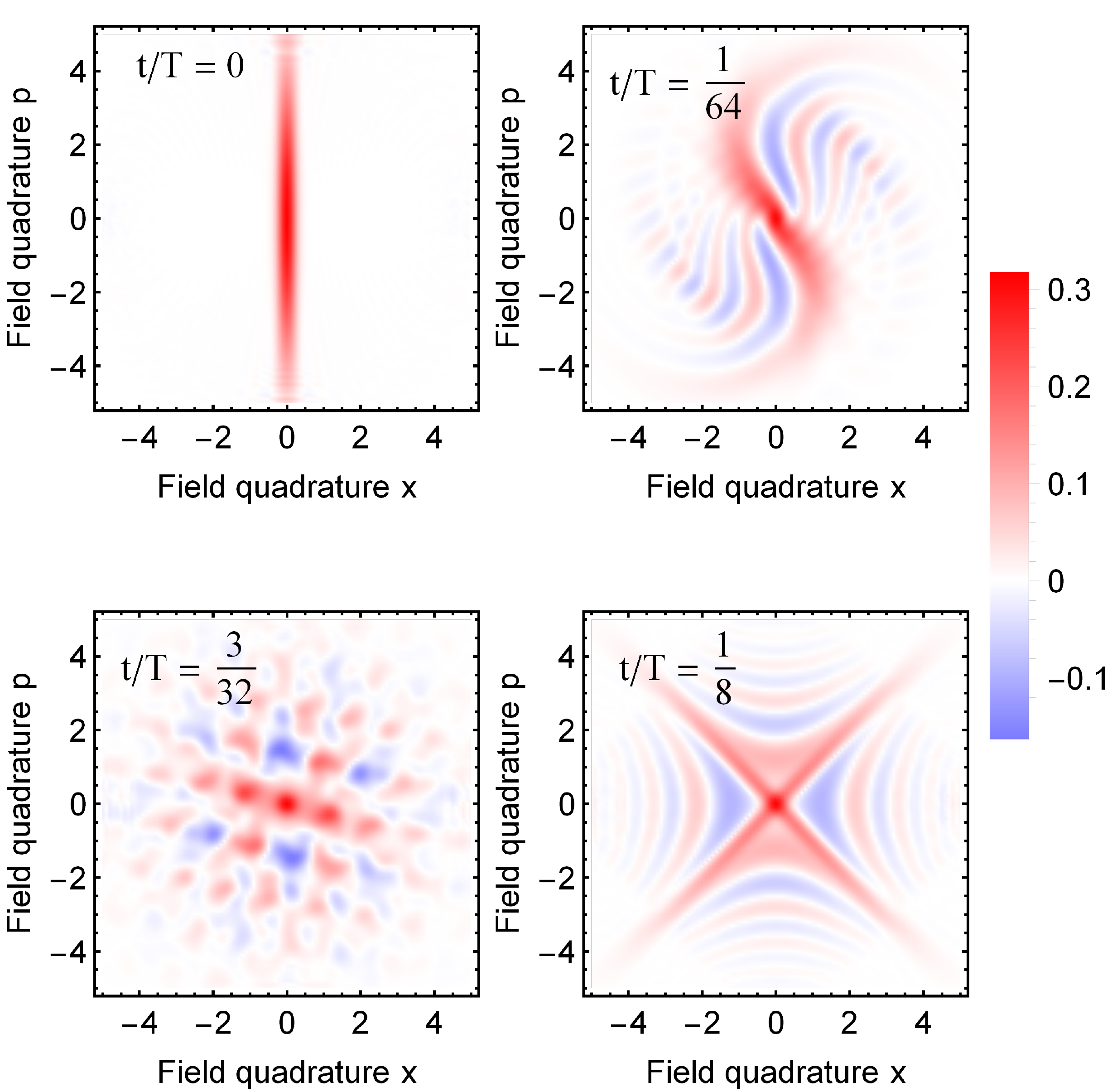}
\caption{The Wigner function for initially squeezed vacuum field at different times of nonlinear evolution with $g/\omega=0.12$.}
\label{fig5}
\end{figure}

The non-Gaussian features of the field can be proved by calculating the Wigner function which is presented in  figure \ref{fig5} in case of initial squeezed vacuum field state at different instants of time. 
The first distribution (at time $t/T=0$) corresponds to the initial state strongly squeezed in $x$ direction. The second one (at time $t/T=1/64$) shows how this distribution is modified at the very beginning of the self-phase modulation process. And then the Wigner function is significantly modified. Many areas of negativity of the Wigner function are seen at time $t/T=3/32$ and $t/T=1/8$ which is a clear evidence of formation of non-Gaussian field state. We have found that at the time $t/T=1/8$ the coherent superposition of two squeezed vacuum states with relative phase $\pi/2$ is formed and can be written as
\begin{equation}\label{cross}
     |\psi(T/8)\rangle= \frac{e^{-i\pi/4}|R\rangle_{\pi/4}+e^{i\pi/4}|R\rangle_{-\pi/4}}{\sqrt2},
\end{equation}
similar to the case of discussed coherent state.

The size of the regions of negativity turns out to be comparable to the size of the positive regions making such state of field relatively robust against losses in the system.  Thus, using the self-phase modulation process it is possible to produce different superposition of coherent and squeezed states with a certain relative phase between them. 

\subsection{Strong coupling regime of interaction}

Now let us consider the case when the excitation of quantum dot becomes important. For coherent initial field state the regimes of collapses and revivals presented in figute \ref{fig1}(a) are found also to take place in the dynamics of the mean number of photons and quadrature vaviance of the quantum field. Such behavior results from the conservation of the mean sum of the total number of photons and excitations in the system, since the operator of this total number appears to be the integral of motion for the Hamiltonian (\ref{ham}). It should be noticed that during the interaction and strong mutual influence between the quantum dot and the field, generally, the total system can not be factorized and each subsystem appears to be in a mixed state.

Such feature is referred to as the entanglement \cite{Fedorov2006,Balybin2018,Popolitova2020} between the quantum dot and the field and its degree can be characterized by the Schmidt parameter which is determined by the inverse trace of the square of reduced quantum dot density matrix $\rho^{qd}$ and can be calculated as follows  \cite{Popolitova2019,Fedorov2006,Popolitova2020}:

\begin{equation}\label{schmidt}
K=\frac{1}{\rho_{11}^{qd}+\rho_{22}^{qd}+2\left|\rho_{12}^{qd}\right|^2}.
\end{equation}

\begin{figure}
\centering
  \includegraphics[width=\linewidth]{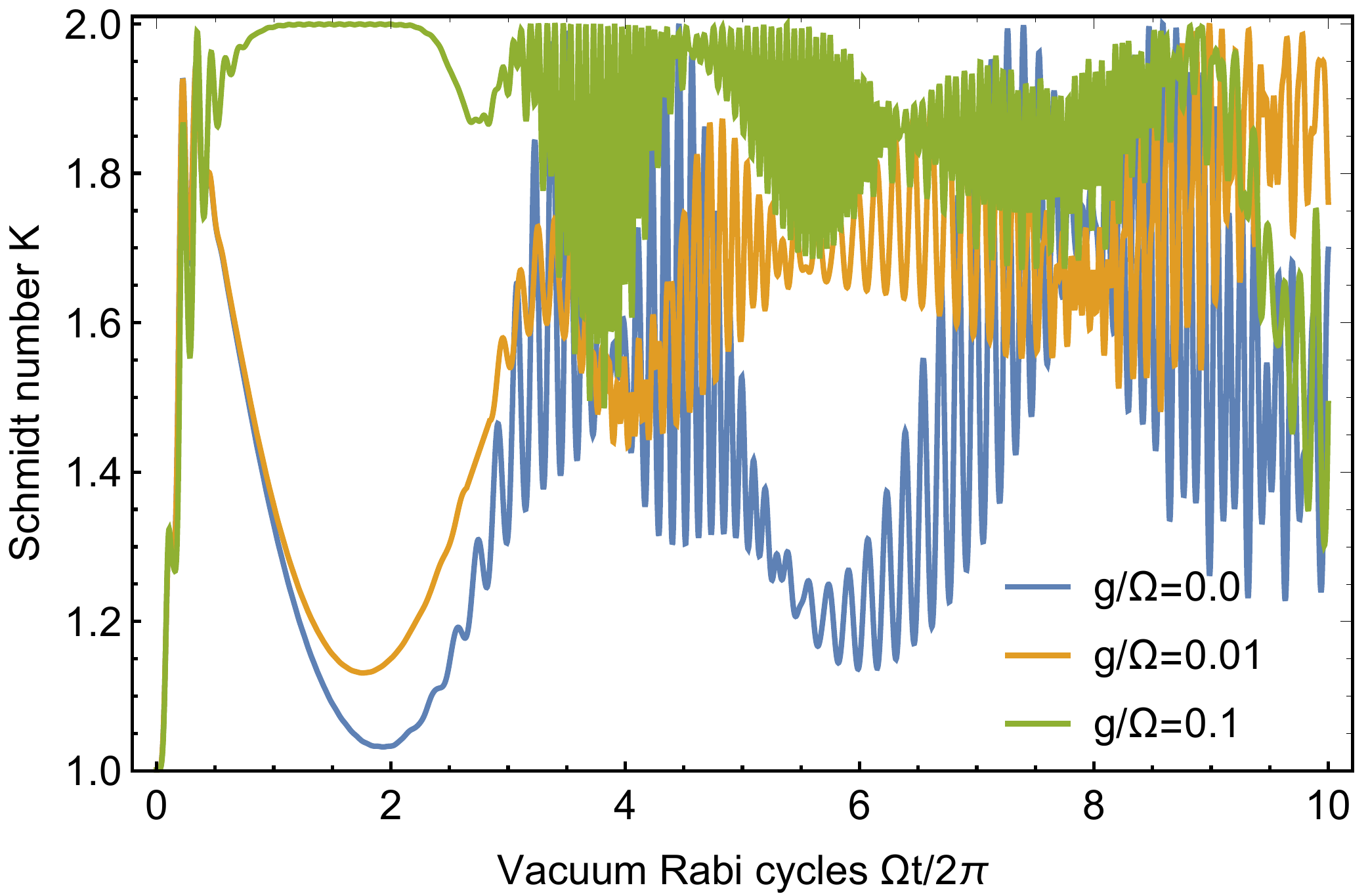}
  \caption{The time-dependent Schmidt parameter $K$ (\ref{schmidt}) calculated for different ratios $g/\Omega$ = 0. 0.01, 0.1.}
  \label{fig6}
\end{figure}

This value can be easily obtained using (\ref{sol_form}-\ref{sol}) and is plotted in dependence on time in figure \ref{fig6} for different values of nonlinearity parameter $g$.

In our case the values of the Schmidt parameter change in the range from 1 to 2 with maximal entanglement being reached for $K=2$. The case $K=1$ corresponds to fully disentangled systems. It is seen that in a strong coupling regime the growing nonlinearity leads to enhancement of entanglement. However, if the nonlinearity begins to prevail, the entanglement goes down and the regime discussed in the previous section takes place.

The comparison between figure \ref{fig6} and figure \ref{fig1}  evidently shows that during the collapse regime when both the excitation probability and the mean number of photons in the field are fixed the dynamics of the system does not stop and manifests itself in the significant change of entanglement. From (\ref{schmidt}) it can be easily seen that during the collapse regime both the quantum dot and field states are changing dramatically since efficient dynamics of the off-diagonal matrix elements for each subsystem takes place.

\begin{figure}
\centering
  \includegraphics[width=\linewidth]{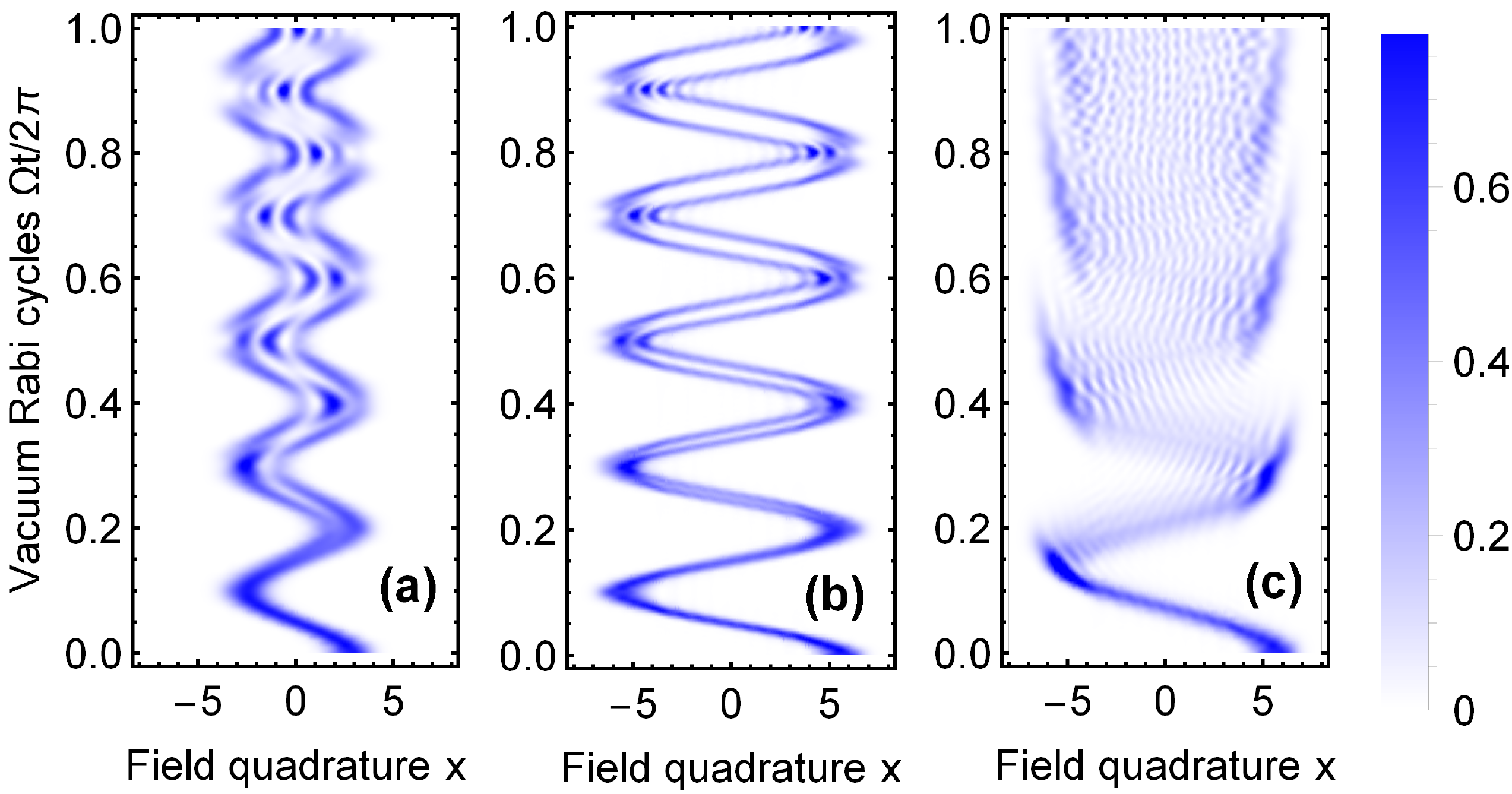}
  \caption{Quantum carpet obtained for coherent initial field state with $\alpha=2$ (a) and $\alpha=4$ (b,c) for $g/\Omega = 0$ (a,b) and $g/\Omega = 0.1$ (c)}
  \label{fig7}
\end{figure}

The evolution of the quantum field state during the interaction with quantum dot under negligibly small nonlinearity is illustrated by quantum carpet presented in figure \ref{fig7}(a,b) for different initial coherent field states.

This data of show firstly almost free dynamics of the field wave packet. Then two close trajectories seem to appear and narrow local maxima of the field density distribution are formed. The formation of two close trajectories corresponds to the double-peak structure of the field wave packet, and can be easily understood from the obtained analytical solution (\ref{sol_form}-\ref{sol}). For $g=0$ the transition amplitudes $a_n$ and $b_n$ found in the form of $\cos(\Omega \sqrt n t)$ and $\sin(\sqrt{n+1} \Omega  t)$ respectively are actually responsible for adding a specific phase to each Fock state and the phase shift can be either positive or negative. For rather high value of $\alpha$ the initial distribution over photon Fock states is relatively narrow and this effect leads to formation of two coherent states with acquired opposite phase shifts. In this case the field density distribution is characterized by double peak structure with displacement of peaks being rather small and depending on the interaction strength. To prove this fact we choose a certain instant of time on figure \ref{fig6} $\Omega t^* / 2 \pi=2$ when the Schmidt parameter is close to unity for $g=0$ and the wave function of the system is almost factorized:

\begin{equation}
\psi(t^*) \approx \frac{|g\rangle-i |e\rangle}{\sqrt{2}} \psi_{field}(t^*).
\end{equation}

\begin{figure}
\centering
  \includegraphics[width=\linewidth]{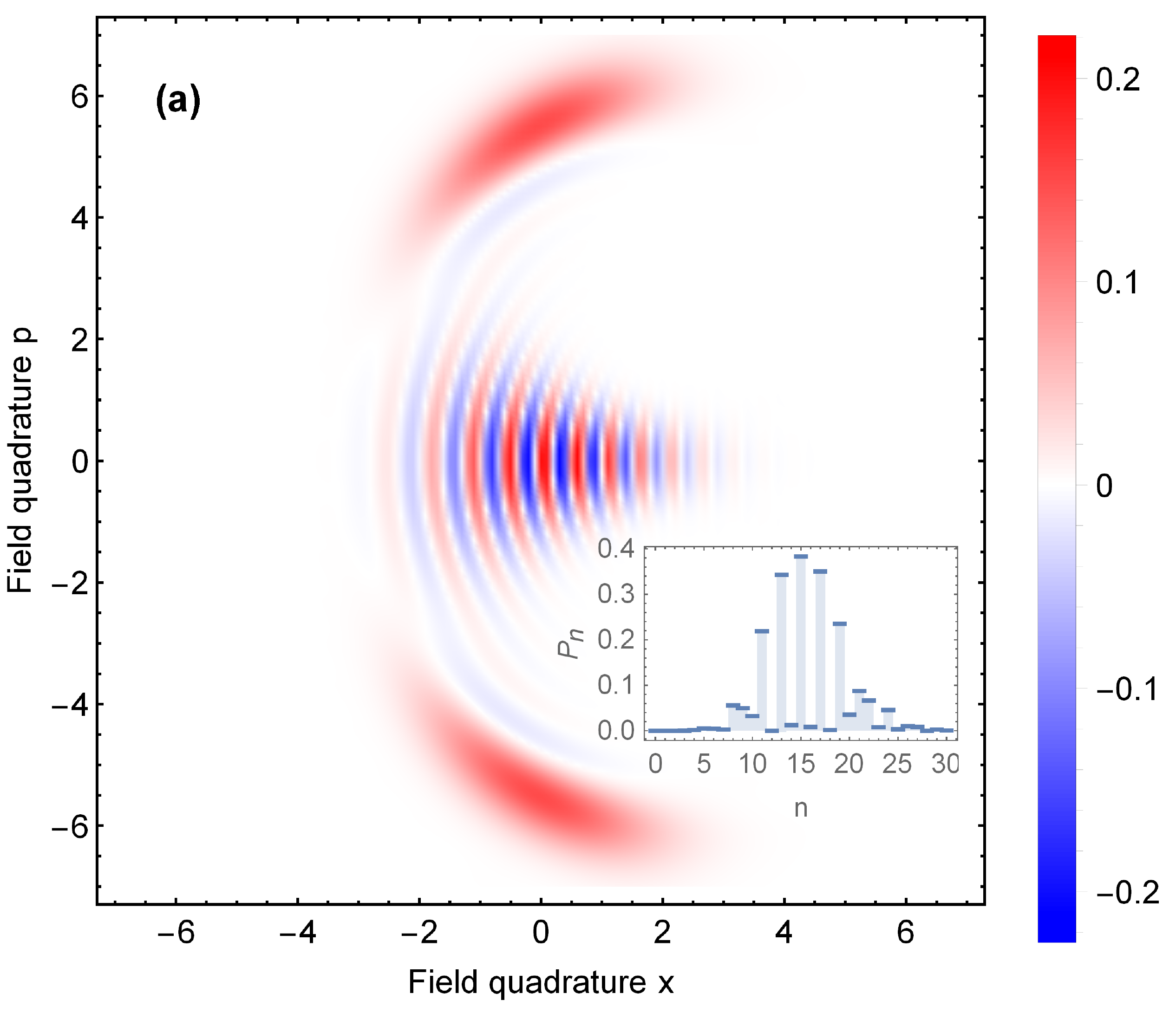}
  \includegraphics[width=\linewidth]{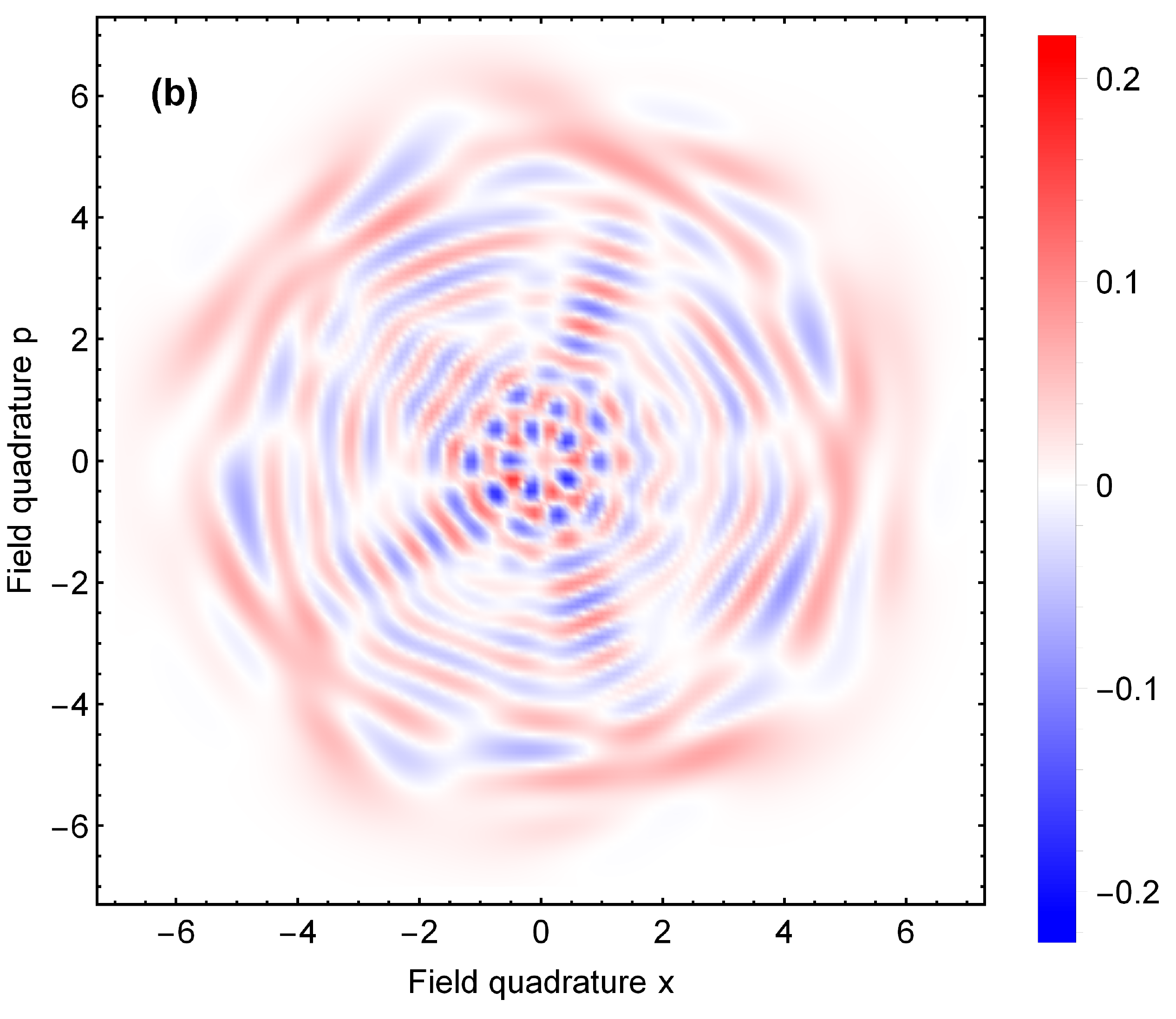}
  \caption{The Wigner function obtained for the pure field state at $g/\Omega=0$ (a) and mixed field state for  $g/\Omega=0.1$  at $\Omega t^*/ 2\pi = 2$.}
  \label{fig8}
\end{figure}

Here we give an explicit form of the quantum dot state found analytically. The features of the field state are demonstrated by the calculated Wigner function presented in figure \ref{fig8}(a).

It is evidently seen that the obtained field state appears to be a superposition of two coherent states with opposite but rather small additional phases such as phase difference is significantly smaller than $\pi/2$. Such state can be referred to a type of “Cat states” and the photon statistics illustrated on the inset of figure \ref{fig8}(a) demonstrates the predominant population of odd Fock states.
During the evolution the presented two peaks on the Wigner function can periodically contribute constructively to the field density distribution and give rise to the bright narrow local maxima observed on the quantum carpet in figure \ref{fig7}(a,b).

At the same time the found state is characterized by interference fringes with negative regions of the Wigner function typical for “cat” states (see figure \ref{fig8}(a)). Thus, the interaction with quantum dot is found to provide directly the formation of non-Gaussian field state.

In the case of non-zero nonlinearity the entanglement between the semiconductor and field subsystems can be very high and most of the time can be equal to its maximum value (see figure \ref{fig6} the curve for $g/\Omega=0.1$). In this case the field appears to be in a mixed state and the field wave packet is highly structured , though local maxima are firstly formed and pronounced on the carpet during the beginning of the evolution (see figure \ref{fig7}(c)). The correspondent Wigner function characterizing now this mixed field state and calculated for the same time moment $t^*$ (maximum of entanglement) is presented in figure \ref{fig8}(b). It is clearly seen that the formed state is significantly non-Gaussian one and of new type which can not be generated simply by self-phase modulation or during the interaction with quantum dot only. Thus the action of a quantum dot in combination with self-phase modulation seems to be a way for generation of different types of non-Gaussian field states.

\begin{figure}
\centering
  \includegraphics[width=\linewidth]{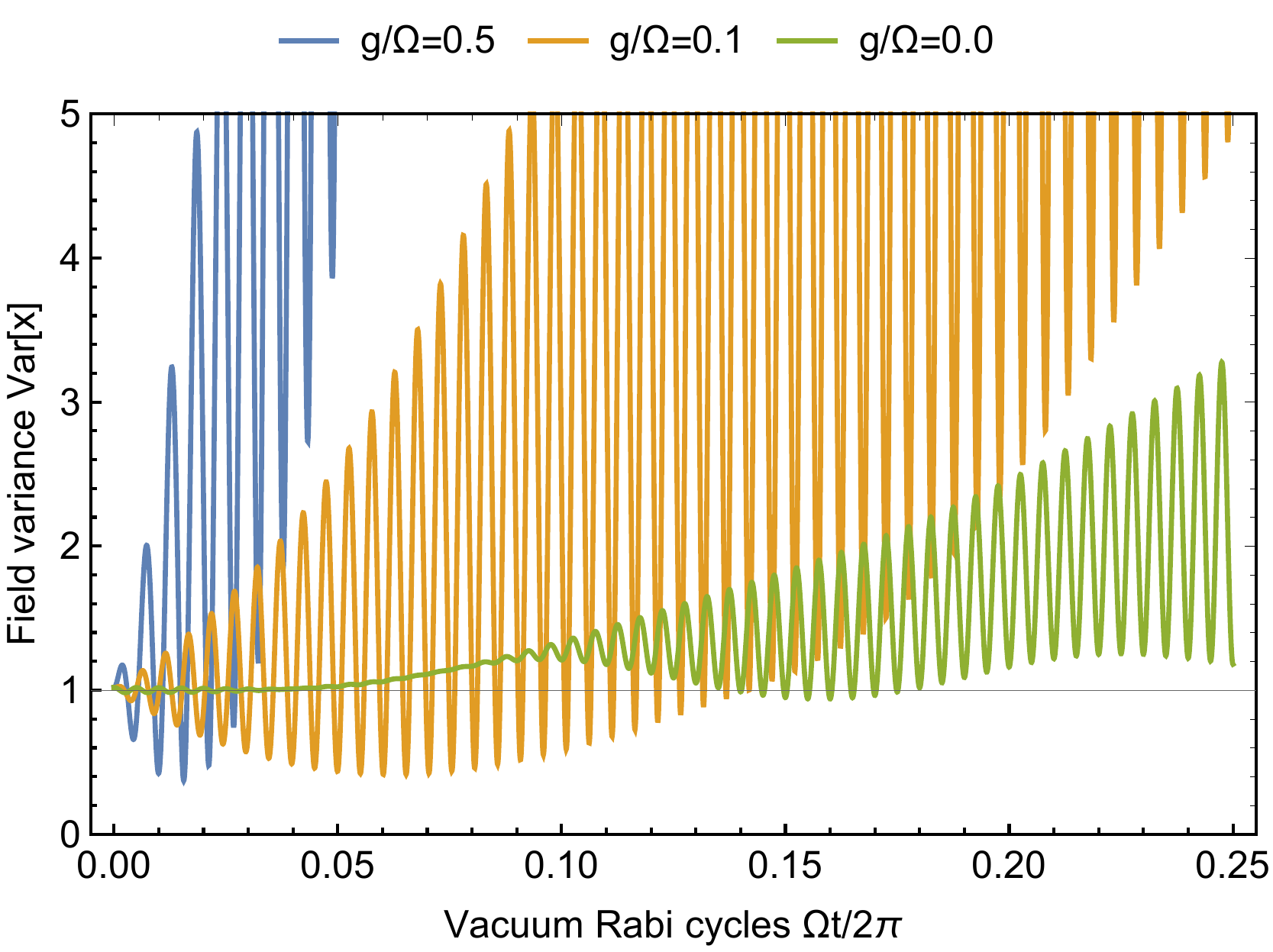}
  \caption{The time-dependent quadrature variance calculated for initial coherent field state with $\alpha=4$ and normalized by the shot noise limit, $\omega/\Omega=100$.}
  \label{fig9}
\end{figure}

In addition we would like to stress that strong coupling with quantum dot accompanied by self-phase modulation provides quadrature squeezing of the interacting field. The time-dependent quadrature squeezing calculated for different ratios $g/\Omega$ is presented in figure \ref{fig9}.

The obtained results demonstrate significant decrease of the quadrature variance below the shot noise level $1/2$. It is important that such squeezing is observed at small times and will be further partly repeated at the revival. Moreover, there is an optimal degree of nonlinearity which provides the highest squeezing. It is determined by the condition when coupling with quantum dot and the Kerr-nonlinearity are mostly equally contributing to the dynamics of the bipartite system.

\section{Conclusion}

In conclusion, we investigate the interaction of a semiconductor quantum dot with a single photon mode in a solid nonlinear cavity and examine the influence of the self-phase modulation effect on the dynamics of bipartite system and quantum dot excitation. The nonlinear phase modulation is found to bring the field out of resonance with the quantum dot transition and provide the suppression of the induced excitation. At the same time, the nonlinearity leads to significant growing of entanglement between the field and semiconductor subsystems if the coupling between them remains strong. The evolution of this entanglement gives the possibility to distinguish different regimes of the quantum dot dynamics as well as to analyze the physics of these regimes by studying the time-dependent behavior of the quantum field subsystem.

The time evolution of electromagnetic field state is examined and significant squeezing of the field quadrature induced by interaction with quantum dot in the presence of nonlinearity is found. The peculiarities of the field dynamics are analyzed using the 2D quantum carpets and the non-Gaussian features of quantum field acquired during the interaction are demonstrated. The non-Gaussian states formed at several instants of time are found explicitly and the possibility to generate several superpositions of coherent or squeezed vacuum states with a certain relative phase is shown. The action of a quantum dot in combination with the self-phase modulation is demonstrated to provide the formation of different and new types of non-Gaussian field states.

\section{Acknowledgment}
We acknowledge financial support of the Russian Science foundation project N\textsuperscript{\underline{o}} 19-42-04105. SNB acknowledges personal support from the Foundation for the Advancement of Theoretical Physics and Mathematics “BASIS”.












%
%

\end{document}